\documentclass[aip,jcp,twocolumn,showpacs,superscriptaddress,amsmath,amssymb,floatfix,10pt,reprint]{revtex4-1}
\usepackage[T1]{fontenc}
\usepackage{graphicx}  
\usepackage{xcolor}
\usepackage{dcolumn}   
\usepackage{bm}        
\usepackage{amssymb}   
\usepackage{framed}
\usepackage{amsmath}
\usepackage{hhline}
\usepackage{color}
\usepackage{multirow}
\definecolor{bluegreen}{rgb}{0,0.2,0.8}
\usepackage{subfigure,amsmath,verbatim,moreverb}
\usepackage{tabularx}
\usepackage{adjustbox}
\usepackage{lipsum}
\usepackage{longtable}
\usepackage{booktabs}
\usepackage{latexsym}
\usepackage{dcolumn}
\usepackage{amsmath}
\usepackage{epsf}
\usepackage{float}
\usepackage{hyperref}

\usepackage{longtable}
\usepackage{booktabs}
\usepackage{threeparttable}

\newcommand{\Eq}[1]{Eq.~\eqref{#1}}
\newcommand{\be}{\begin{equation}}
\newcommand{\ee}{\end{equation}}
\newcommand{\bea}{\begin{eqnarray}}
\newcommand{\eea}{\end{eqnarray}}
\newcommand{\bean}{\begin{eqnarray*}}
\newcommand{\eean}{\end{eqnarray*}}

\usepackage{tikz}
\usetikzlibrary{arrows.meta, positioning}

\usepackage{etoolbox}
\AtBeginEnvironment{align}{\setcounter{subeqn}{0}}
\newcounter{subeqn} %

\begin{document}


\title[]{
Accurate starting points for one-shot G$_0$W$_0$ and Bethe-Salpeter Equation calculations via effective tuning of range-separated hybrid functionals}
\author{Aditi Singh}
\email{aditisingh4812@gmail.com}
\affiliation{Institute of Physics, Faculty of Physics, Astronomy and Informatics, Nicolaus Copernicus University in Toru\'n,
ul. Grudzi\k{a}dzka 5, 87-100 Toru\'n, Poland}
 \affiliation{Institute of Advanced Studies, Nicolaus Copernicus University in Toru\'{n}, ul. Wile\'{n}ska 4, 87-100 Toru\'{n}, Poland}

\author{Subrata Jana}
 \affiliation{Institute of Physics, Faculty of Physics, Astronomy and Informatics, Nicolaus Copernicus University in Toru\'n,
ul. Grudzi\k{a}dzka 5, 87-100 Toru\'n, Poland}
 \affiliation{Institute of Advanced Studies, Nicolaus Copernicus University in Toru\'{n}, ul. Wile\'{n}ska 4, 87-100 Toru\'{n}, Poland} 

\author{Szymon \'Smiga}
\email{szsmiga@fizyka.umk.pl}
\affiliation{Institute of Physics, Faculty of Physics, Astronomy and Informatics, Nicolaus Copernicus University in Toru\'n,
ul. Grudzi\k{a}dzka 5, 87-100 Toru\'n, Poland}
 \affiliation{Institute of Advanced Studies, Nicolaus Copernicus University in Toru\'{n}, ul. Wile\'{n}ska 4, 87-100 Toru\'{n}, Poland}

\date{\today}

\begin{abstract}
The accuracy of one-shot G$_0$W$_0$ and Bethe-Salpeter equation (BSE) calculations depends strongly on the underlying starting-point eigensystem, which is commonly obtained from a mean-field density-functional approximation (DFA). Range-separated hybrid (RSH) functionals provide a particularly effective starting point, however, conventional optimally tuned RSH procedures often require costly, system-specific, multi-step optimizations of the range-separation parameter $\mathrm{\omega}$. In this work, we show that a recently proposed effective tuning protocol [Singh \textit{et. al.}, Journal of Physical Chemistry Letters, 16, 32, 8198-8208, (2025)] for RSH functionals can serve as an efficient alternative for determining $\mathrm{\omega}$ used in G$_0$W$_0$ and BSE calculations. This simplified tuning scheme yields range-separation parameters that are effectively equivalent to those obtained from more elaborate tuning strategies, while avoiding their substantial computational overhead. The resulting tuned RSH eigensystems provide reliable starting points for many-body perturbation theory (MBPT). In particular, one-shot G$_0$W$_0$ calculations based on effectively tuned RSH orbitals reproduce reference ionization potentials (IPs) with high accuracy, while subsequent BSE calculations yield quantitatively reliable neutral excitation energies, optical absorption spectra, and excitonic properties for a diverse set of molecular systems and clusters. These results demonstrate that effective RSH tuning offers a practical and broadly applicable route to accurate quasiparticle and excited-state calculations, combining the accuracy of optimally tuned starting points with the low computational cost required for routine applications of G$_0$W$_0$ and BSE.
\end{abstract}

\maketitle

\section{Introduction}

Many-body perturbation theory (MBPT) methods, particularly GW and the Bethe-Salpeter equation (BSE), are routinely applied to predict properties of molecules, clusters, and materials with high precision~\cite{Onida2002electronic,Leng2016GW,Reining2018GW,Louie2005book,gw-review,Blase2020BSE,Bruneval2021GW}. This progress has been driven by advances in software~\cite{berkeleygw,fhiaimsgw1,fhiaimsgw2,Ren2012RI,Caruso2013self,molgw-code,nwchemgw,nwchem,Neuhauser2014Breaking}, scalable algorithms~\cite{Rohlfing2000Electronhole,Liu2016Cubic,Albrecht1998Abinitio,Hirose2015All,Cho2022Simplified,Marsili2021Spinorial,Dvorak2019Quantum,Brian2024manybody,Marsili2017largescale,Sharma2024Optical,Barrueta2023Accelerating,Duchemin2021Cubic,Villalobos2023Lagragian,Gao2024Efficient,SwiftGW,Vlcek2017Stochastic,Wen2024Comparing}, and high-performance computing~\cite{Ben2020Accelerating}.

In the GW approximation, many-body effects enter via self-energy corrections to Kohn-Sham (KS) states~\cite{Blase2020BSE}. The exact self-energy requires solving coupled equations self-consistently~\cite{Blase2020BSE,Blase2018}, but in practice $\Sigma$ is often approximated to first order in G and W~\cite{gw-review}, with G constructed from a density-functional approximation (DFA). Different self-consistency levels exist; fully self-consistent GW (scGW) is rigorous but computationally demanding. Thus, the one-shot G$_0$W$_0$ method
~\cite{Golze2019GW,Stan2009Levels,Setten2015GW100,Atalla2013Hybrid,Fuchs2007Quasiparticle,Huser2013Quasiparticle} is widely used for its favorable accuracy-cost balance.

G$_0$W$_0$ systematically improves the description of quasiparticle electronic structure over standard KS density functional theory (DFT)~\cite{burke2012perspective}, yielding accurate ionization potentials (IPs)~\cite{ortiz-24-ip,Bruneval2013Benchmarking,ot-rsh-gw100,Tolle2023} and fundamental band gaps~\cite{Hybertsen1885First, Hybertsen1986electron,Zhu2021, Jana2025nonempirical,Ghosh2025Advancing,Ghosh2026Accurate,Tolle2023}. Moreover, the resulting quasiparticle energies provide an accurate starting point for the Bethe-Salpeter equation (BSE), enabling reliable predictions of neutral excitation energies~\cite{Ghosh2026Accurate,bse-thiel,ot-rsh-gw100} and exciton binding energies~\cite{Ghosh2025Advancing,Bhattacharya2024}. These improvements arise from the nonlocal and energy-dependent self-energy, which provides a more realistic description of electron addition and removal processes than the approximate exchange-correlation (XC) potentials employed in conventional DFT~\cite{Hedin1965GW,Blase2018,gw-review,Onida2002electronic}.\\
Nevertheless, G$_0$W$_0$ accuracy depends strongly on the initial starting point~\cite{Setten2015GW100,Tonatiuh2020Reproducibility,Gant2022optimally,Zhang2022Recommendation,Bruneval2013Benchmarking}, typically obtained from semilocal or hybrid functionals~\cite{Hedin1999correlation,Tiago2004Effect,Louie1987Theory,Setten2015GW100,Li2005Quasiparticle,Miyake2006Quasiparticle,Bruneval2006Effect,Kioupakis2008GW,Sette2017Automation,Palummo2009abinitio,Jiang2009Localized,Marom2012Benchmark,Atalla2013Hybrid,Duchemin2016Combining,Refaely2018Defect,Wilhelm2021LowScaling,Golze2019GW,Fuchs2007Quasiparticle,Salas2022Electronic,Shishkin2007Selfconsistent,Shishkin2007accurate,Huser2013Quasiparticle}. Semilocal functionals are often less favorable~\cite{Rinke_2005,Rinke_2006,Zhang2022Recommendation,Bruneval2013Benchmarking,Gant2022optimally} due to their well-known deficiencies, including self-interaction and delocalization errors and the absence of the correct asymptotic exchange potential~\cite{PhysRevLett.51.1884,gw-review,PhysRevLett.51.1888,Mori2008localization,Zhang2022Recommendation,Setten2015GW100}. Range-separated hybrid (RSH) DFAs partially remedy these shortcomings~\cite{heyd2003hybrid,krukau2006influence,heyd2004efficient,jana2020screened,jana2020improved,jana2018efficient,janameta2018,jana2019screened} by combining short-range DFT exchange with long-range Hartree-Fock exchange, thereby restoring the correct $-1/r$ asymptotic potential~\cite{Potential-Gilbert,Potential-issue,JPCLVignesh}.\\
The use of optimally tuned range-separated hybrid (OTRSH) functionals as starting points for MBPT has become a well-established strategy over the past decade. Refaely-Abramson \textit{et al.} first suggested that enforcing the IP theorem through optimal tuning (OT) provides an accurate starting point for molecular GW calculations~\cite{Refaely2012Quasiparticle}. This approach was subsequently shown to significantly improve molecular quasiparticle energies within GW by Gallandi and Körzdörfer~\cite{Gallandi2015} and later extended to molecular GW-BSE calculations by McKeon \textit{et al.}~\cite{ot-rsh-gw100}. Similar advances have also been achieved for periodic systems, where OTRSH starting points substantially improve both GW quasiparticle energies and GW-BSE excitation spectra~\cite{Gant2022optimally,Ohad2023}. These successes arise from the ability of OTRSH functionals to satisfy physically motivated constraints on frontier orbital energies while providing the correct asymptotic exchange potential. Consequently, OTRSH functionals have been shown to yield accurate IPs from the HOMO energy~\cite{SteiEiseHele2010,Tamblyn2014,Kanchanakungwankul2021Examination,Baer2010Tuned}, improved HOMO-LUMO gaps and fundamental band gaps~\cite{SteiEiseHele2010,Tamblyn2014,Wing2021band,OhadWingGant2022,WeiGiaRigPas2018,jana2023simple,Ghosh2024accurate}, a reliable description of charge-transfer (CT) excitations~\cite{Stein2009Reliable,Baer2010Tuned,Kronik2012excitation,kronik2018dielectric}, and accurate time-dependent density functional theory (TDDFT) excitation energies~\cite{Stein2009Reliable,Kronik2012excitation,kronik2018dielectric,Knight2016Accurate,Richard2016Accurate,singh2025simplifiedphysicallymotivateduniversally}. Despite these demonstrated advantages, determining the optimal range-separation parameter requires repeated generalized Kohn-Sham (GKS) self-consistent-field (SCF) calculations for neutral and ionic species, making the conventional OT procedure computationally demanding, particularly for large systems, high-throughput screening, and applications requiring calculations on many related structures~\cite{kronik2018dielectric,Karolewski2013Using,Refaely2012Quasiparticle}. These challenges motivate the development of simplified and more transferable tuning strategies that retain the accuracy of OTRSH approaches while broadening their applicability to a wider range of functionals.
A recent promising development is a simplified effective tuning scheme ($\omega_{\mathrm{eff}}$) that avoids multiple SCF calculations~\cite{singh2025simplifiedphysicallymotivateduniversally}. This density-dependent scheme performs well even for CT excitations, open-shell radicals, clusters, layered materials, and bulk solids.\\
Motivated by these advances, we demonstrate in this work that the $\omega_{\mathrm{eff}}$ tuning scheme provides a practical, black-box starting point for G$_0$W$_0$ and BSE calculations, eliminating the need for conventional system-by-system OT. This approach is inexpensive, broadly applicable, and retains accuracy for charged and neutral excitations. We benchmark IPs and excitation energies against established datasets. Additionally, given the importance of optical gaps in quantum-confined systems~\cite{PRB1993Si, PRL1999Si, PRL2000Si}, we also assess the method for hydrogenated silicon quantum dots.\\
The paper is organized as follows: Sec.~\ref{sec_method} presents the theoretical background and computational methodology; Sec .~\ref{sec:results} presents results and discussion; and Sec.~\ref{sec:conclusion} concludes with future perspectives.

\section{Theoretical Background and Computational Details}
\label{sec_method}

\subsection{Tuning Approach for RSH}

To investigate the sensitivity of G$_0$W$_0$ and BSE results to the choice of tuning scheme, we employ the LC-$\mathrm{\omega}$PBEh RSH functional~\cite{lcwpbeh}. This functional combines the PBE semilocal density functional approximation with fixed exchange parameters $\alpha=0.20$ and $\beta=0.80$~\cite{camb3lyp}. The choice of $\alpha$ is motivated by previous studies, which identified values in the range $0.20$--$0.25$ as optimal for molecular systems~\cite{Refaely2012Quasiparticle,Kronik2012excitation}. LC-$\mathrm{\omega}$PBEh was selected because it has been shown to provide accurate starting points for MBPT calculations, in particular for G$_0$W$_0$ and BSE calculations~\cite{ot-rsh-gw100}. Its relatively large fraction of exact exchange is also important for improving IP predictions in both KS/GKS~\cite{JPCLVignesh} and G$_0$W$_0$ calculations~\cite{Bruneval2013Benchmarking}.

We compare three schemes for determining the range-separation parameter $\omega$:
\begin{equation}
\label{eq:omega_form}
\begin{cases}
\omega_{\mathrm{OTRSH}} =
\arg\min_{\omega}
\left[
\begin{aligned}
&\left| \mathrm{IP}(\omega) + \varepsilon_{\mathrm{HOMO}}(\omega) \right|^2 \\
&+ \left|\mathrm{EA}(\omega) + \varepsilon_{\mathrm{LUMO}}(\omega) \right|^2
\end{aligned}
\right],
\\[0.8em]

\omega_{\mathrm{HOMO}} =
\arg\min_{\omega}
\left|
\varepsilon_{\mathrm{HOMO}}^{G_0W_0@\mathrm{GKS}(\omega)}
-
\varepsilon_{\mathrm{HOMO}}^{\mathrm{GKS}(\omega)}
\right|,
\\[0.8em]

\omega_{\mathrm{eff}} =
\dfrac{a_1}{\langle r_s \rangle}
+
\dfrac{a_2 \langle r_s \rangle}
{1 + a_3 \langle r_s \rangle^2}.
\end{cases}
\end{equation}

The optimally tuned RSH scheme, denoted here as OTRSH, determines $\mathrm{\omega}$ by enforcing Koopmans'-like conditions for the frontier orbitals~\cite{Kronik2012excitation,Refaely2012Quasiparticle}. Specifically, the HOMO energy is tuned to reproduce $-\mathrm{IP}$, while the LUMO energy is tuned to reproduce $-\mathrm{EA}$, where
$\mathrm{IP} = \mathrm{E(N-1) - E(N)}$ and $\mathrm{EA} = \mathrm{E(N) - E(N+1)}$, respectively. This procedure also aligns the HOMO-LUMO gap with the fundamental gap, $\mathrm{IP} - \mathrm{EA}$. When the anionic state is unbound, the tuning is commonly restricted to the IP condition only.

In addition to OTRSH tuning, we consider the HOMO-tuning scheme~\cite{Atalla2013Hybrid,Rangel2017assessment,ot-rsh-gw100}. In this approach, the optimal value $\omega_{\mathrm{HOMO}}$ is obtained by matching the HOMO QP energy from a G$_0$W$_0$ calculation performed on top of a GKS reference quantities 
to the corresponding HOMO eigenvalue of the same GKS calculation. This condition enforces a generalized Koopmans' theorem and reduces the dependence of the quasiparticle energy on the starting-point functional.

The third scheme employs an effective range-separation parameter,
$\omega_{\mathrm{eff}}$, obtained using a physically motivated effective tuning
scheme developed in our previous work~\cite{singh2025simplifiedphysicallymotivateduniversally}.
The underlying idea is that the range-separation parameter can be related to an
effective screening parameter through the long-wavelength limit of the XC kernel of the uniform electron gas. By combining the
compressibility sum rule with the local-density approximation, an analytical
relationship between the screening parameter and the density-weighted
Wigner-Seitz (WS) radius is obtained. For computational efficiency, this
theoretically derived relationship is represented by a simple analytical
expression, from which the effective range-separation parameter is determined.\\
Specifically, $\omega_{\mathrm{eff}}$ is given by the final expression in
Eq.~\eqref{eq:omega_form}, where the coefficients
$a_1=1.91718$, $a_2=-0.02817$, and $a_3=0.14954$
follow from the analytical representation of the theoretically derived
relationship presented in Ref.~\cite{singh2025simplifiedphysicallymotivateduniversally}. The key quantity entering this expression is the density-weighted average WS radius $(\langle r_s\rangle)$, which characterizes the effective electronic environment of the system and is defined as
\begin{equation}
\langle r_{s} \rangle =
\frac{
\displaystyle\int
\mathrm{erf}\left(\frac{n(\mathbf{r})}{n_{\mathrm{c}}}\right)
r_s(\mathbf{r}) d^3\mathbf{r}
}{
\displaystyle\int
\mathrm{erf}\left(\frac{n(\mathbf{r})}{n_{\mathrm{c}}}\right)
 d^3\mathbf{r}
},
\label{eq:rs-average}
\end{equation}
where $r_s(\mathbf{r})$ is the local WS radius. The error-function weighting smoothly suppresses contributions from
regions where the electron density falls below the cutoff density
$n_{\mathrm{c}}$, effectively excluding the diffuse low-density tail
while retaining the physically relevant electron-density region. The
cutoff density $n_{\mathrm{c}}$ is defined as
\begin{equation}
n_{\mathrm{c}}=
\frac{n_{\mathrm{th}}}
{\displaystyle\int n(\mathbf{r})\,d^3\mathbf{r}},
\label{Eq:error2}
\end{equation}
where the denominator is the total number of electrons, $N$, in the
system. Consequently, $n_{\mathrm{c}}$ is system dependent and
naturally scales with the size of the electronic system. 
Throughout this work, the reference density parameter is fixed at
$n_{\mathrm{th}}=1.64\times10^{-2}\,\mathrm{e/bohr^3}$, where $\mathrm{e}$ denotes the number of electrons, rather than the
elementary charge. A detailed derivation, parameterization, and validation of the effective tuning scheme are provided in Ref.~\cite{singh2025simplifiedphysicallymotivateduniversally}.

A schematic comparison of the $\mathrm{\omega_{OTRSH}}$ and $\mathrm{\omega_{eff}}$ workflows is presented in Fig.~\ref{fig:workflow}.

\subsection{G$_0$W$_0$ and BSE Approach}

From the RSH starting point, quasiparticle (QP) energies are obtained within the one-shot G$_0$W$_0$ approximation~\cite{Hedin1965GW,Hybertsen1986electron}:
\begin{equation}
\varepsilon_n^{\mathrm{QP}} = \varepsilon_n^{\mathrm{RSH}} + \langle \phi_n^{\mathrm{RSH}} | \Sigma(\varepsilon_n^{\mathrm{QP}}) - V_{xc}^{\mathrm{RSH}} | \phi_n^{\mathrm{RSH}} \rangle,
\label{qp_eqn}
\end{equation}
where $\Sigma = \frac{i}{2\pi} \int d\omega' \, \mathrm{G}_0(\omega+\omega') \mathrm{W}_0(\omega')$. IPs are extracted as the negative of the highest occupied QP level. The accuracy of G$_0$W$_0$ depends strongly on the starting point~\cite{Setten2015GW100,Golze2019GW}.

Neutral excitations are obtained from the BSE Hamiltonian~\cite{Onida2002electronic}:
\begin{equation}
\begin{pmatrix}
\mathbf{A} & \mathbf{B} \\
-\mathbf{B}^* & -\mathbf{A}^*
\end{pmatrix}
\begin{pmatrix}
\mathbf{X}^s \\
\mathbf{Y}^s
\end{pmatrix}
=
\Omega_s
\begin{pmatrix}
\mathbf{X}^s \\
\mathbf{Y}^s
\end{pmatrix},
\end{equation}
with matrix elements
\begin{align}
A_{ia,jb} &= \delta_{ij}\delta_{ab}(\varepsilon_a^{\mathrm{QP}} - \varepsilon_i^{\mathrm{QP}})
+ 2\langle ia | V_H | jb \rangle - W_{ijab}, \\
B_{ia,jb} &= 2\langle ia | V_H | bj \rangle - W_{ibaj},
\end{align}
where W is the statically screened Coulomb interaction. The optical absorption spectrum is obtained from the imaginary part of the dielectric function, $\varepsilon_2(\omega)$, whose peaks correspond to optically allowed electronic excitations. Since BSE relies on G$_0$W$_0$ QP energies, the quality of the optical spectra is directly tied to the underlying starting point. Further technical details of the G$_0$W$_0$ and BSE calculations are provided in Supporting Information (SI)~\cite{supplementary}.

\subsection{Computational Details and Test Sets}

Using the methodology described above, we evaluate the robustness and accuracy of the tuning schemes for predicting IPs in the GW100 test set and optical properties in the Thiel set and hydrogenated silicon quantum dots, employing the G$_0$W$_0$ and BSE approaches. \\ 

\textbf{I. The GW100 test set~\cite{Setten2015GW100}:} An IP benchmark comprising 100 small molecules that contains diverse chemical bonds and environments, as it includes different elements. Thus covering a wide range of IPs (from $\sim 4eV$ ($Rb_2$) to $\sim 25eV$ (He)). 
In this case, all calculations are performed using the def2-TZVPP\cite{def2-family} basis set, and effective core potentials (ECPs) \cite{ECP} are utilized for heavy elements. To ensure a rigorous benchmark, we compared our results against high-level CCSD(T) reference IPs~\cite{Klopper2023,Bruneval2021GW} and experimental data from Ref.~\cite{Setten2015GW100}. Because a bound anionic state does not exist for most of these systems, the OTRSH parameter is obtained through IP tuning for all the systems. The $\omega_{\mathrm{OTRSH}}$ (from IP tuning) and $\omega_{\mathrm{HOMO}}$ have been taken from Ref.~\cite{ot-rsh-gw100}. \\

\textbf{II. Thiel test set~\cite{silva2010benchmarks,bte-ref}:}~This comprehensive benchmark provides highly accurate 103 singlet and 63 triplet reference excitation energies for 28 small to medium-sized organic molecules (made of C, N, O, and H elements), including unsaturated aliphatic hydrocarbons, aromatics, heterocycles, carbonyls, amides, and nucleobases. The molecular geometries were taken from Ref.~\cite{bte-ref}. 
The TZVP basis set was uniformly adopted for all calculations to ensure direct comparability with the established coupled cluster benchmarks~\cite{bte-ref}. Although the absence of diffuse functions in this basis set causes convergence problems for the first excitation in ethene and the first bright excitation in pyrrole, this effect is systematic and similarly observed across all computational methods (including CCSD)~\cite{bse-thiel}. The basis set is validated by the convergence analysis with respect to excitation energies reported in Ref.~\cite{bse-thiel}, which we adopt here. The frozen-core approximation was employed throughout. Within this set, only hexatriene, benzoquinone, octatetraene, and tetrazine possess bound anionic states, whereas all other molecules exhibit unbound anionic states. For these unbound systems, the OTRSH parameter is determined via IP tuning. The range-separation parameters, $\omega_{\mathrm{OTRSH}}$, were taken directly from Ref.~\cite{ot-rsh-gw100}. The corresponding $\omega_{\mathrm{HOMO}}$ values were computed in this work using the LC-$\mathrm{\omega}$PBEh functional with the def2-TZVPP basis set, employing the MOLGW code~\cite{molgw-code}. \\

\textbf{III. Hydrogenated Silicon quantum dots (SiQDs):}
These systems are biocompatible, low-toxicity, and abundant materials~\cite{SIQD} with excellent optical properties~\cite{Furukawa-1988,SIQD-1}, making them attractive for applications in biomedicine, catalysis, and optoelectronics~\cite{Takagi1990}. Their strong quantum confinement gives rise to pronounced size-dependent variations in IPs, HOMO-LUMO gaps, and optical excitation energies. These characteristics make SiQDs valuable model systems for investigating how DFT and MBPT describe quasiparticle and excitonic effects in confined semiconductors. Accordingly, they have also served as important benchmark systems for the development and assessment of OTRSH functionals and their ability to predict IPs and fundamental gaps~\cite{SteiEiseHele2010}.
All molecular geometries were optimized in the gas phase using KS-DFT with the B3LYP functional~\cite{becke1993density, lee1988development} in conjunction with the cc-pVDZ basis set~\cite{Dunning}, as implemented in the \textsc{ORCA} software package~\cite{ORCA5}. The optimized structures were verified to be true local minima by confirming the absence of imaginary frequencies in the harmonic vibrational analysis. Cartesian coordinates of all optimized geometries are provided in the public repository~\cite{repo,singh2026_zenodo}. To manage the computational cost, this study focuses on four representative systems (SiH$_4$, Si$_2$H$_6$, Si$_5$H$_{12}$, and Si$_{10}$H$_{16}$) using the TZVP basis set. Additionally, the frozen-core approximation has been employed. Both $\omega_{\mathrm{OTRSH}}$ (via IP tuning) and $\omega_{\mathrm{HOMO}}$ for SiQDs were calculated using the LC-$\mathrm{\omega}$PBEh/def2-TZVPP functional within MOLGW. The corresponding range-separation parameters are provided in the SI~\cite{supplementary}.

\begin{figure}[h!]
\centering
\begin{tikzpicture}[
    node distance=0.7cm and 1.8cm,
    box/.style={
        draw,
        rounded corners,
        very thin,
        align=center,
        minimum width=1.8cm,
        minimum height=0.45cm,
        font=\scriptsize,
        fill=blue!10
    },
    decision/.style={
        draw,
        rounded corners,
        very thin,
        align=center,
        minimum width=2.2cm,
        minimum height=0.55cm,
        font=\scriptsize,
        fill=orange!18
    },
    finalbox/.style={
        draw,
        rounded corners,
        very thin,
        align=center,
        minimum width=2.0cm,
        minimum height=0.45cm,
        font=\scriptsize,
        fill=green!15
    },
    calcbox/.style={
        draw,
        rounded corners,
        very thin,
        align=center,
        minimum width=1.9cm,
        minimum height=0.45cm,
        font=\scriptsize,
        fill=purple!12
    },
    arrow/.style={-{Latex[length=1mm]}, very thin}
]

\node[box] (guess) {$\mathrm{\omega}$ guess};
\node[box, below=of guess] (solve) {Solve RSH};
\node[decision, below=of solve] (check) {Does OTRSH\\satisfy IP condition?};
\node[box, below=of check] (change) {Update $\mathrm{\omega}$};
\node[finalbox, below=of change] (final) {$\omega_{\mathrm{OTRSH}}$};

\node[calcbox, below=of final] (lc) {RSH};
\node[calcbox, below=of lc] (gw) {G$_0$W$_0$};
\node[calcbox, below=of gw] (bse) {BSE};

\node[box, right=1.8cm of check] (pbe) {PBE};
\node[finalbox, below=0.7cm of pbe] (weff) {$\omega_{\mathrm{eff}}$};

\draw[arrow] (guess) -- (solve);
\draw[arrow] (solve) -- (check);

\draw[arrow] (check) -- node[right, font=\scriptsize] {No} (change);

\draw[arrow] (change.west) -- ++(-1.2,0) |- (solve.west);

\draw[arrow] (check.east) -- ++(0.6,0)
    node[midway, above, font=\scriptsize] {Yes}
    |- (final.east);

\draw[arrow] (final) -- (lc);
\draw[arrow] (lc) -- node[right, font=\scriptsize] {$\varepsilon_0,\phi_0$} (gw);
\draw[arrow] (gw) -- (bse);

\draw[arrow] (pbe) -- node[right, font=\scriptsize] {density} (weff);
\draw[arrow] (weff.south) |- (lc.east);

\end{tikzpicture}
\caption{Workflow for $\mathrm{\omega}$ tuning using the OTRSH and $\omega_{\mathrm{eff}}$ schemes, followed by RSH, G$_0$W$_0$, and BSE calculations.}
\label{fig:workflow}
\end{figure}

All G$_0$W$_0$ and BSE calculations were performed using the \textsc{MOLGW} software package~\cite{molgw-code}. For technical details, check the SI~\cite{supplementary}.
Finally, the $\omega_{\mathrm{eff}}$ parameters for all studied systems were obtained using PySCF~\cite{pyscf} with a script available in a public repository~\cite{repo,singh2026_zenodo}. For the GW100 set, $\omega_{\mathrm{eff}}$ was determined using the PBE functional with the def2-QZVP basis set. For the Thiel set and the silicon clusters, $\omega_{\mathrm{eff}}$ was obtained using the PBE functional together with the cc-pVDZ basis set.

\begin{figure}
    \centering   
    \includegraphics[width=\columnwidth]{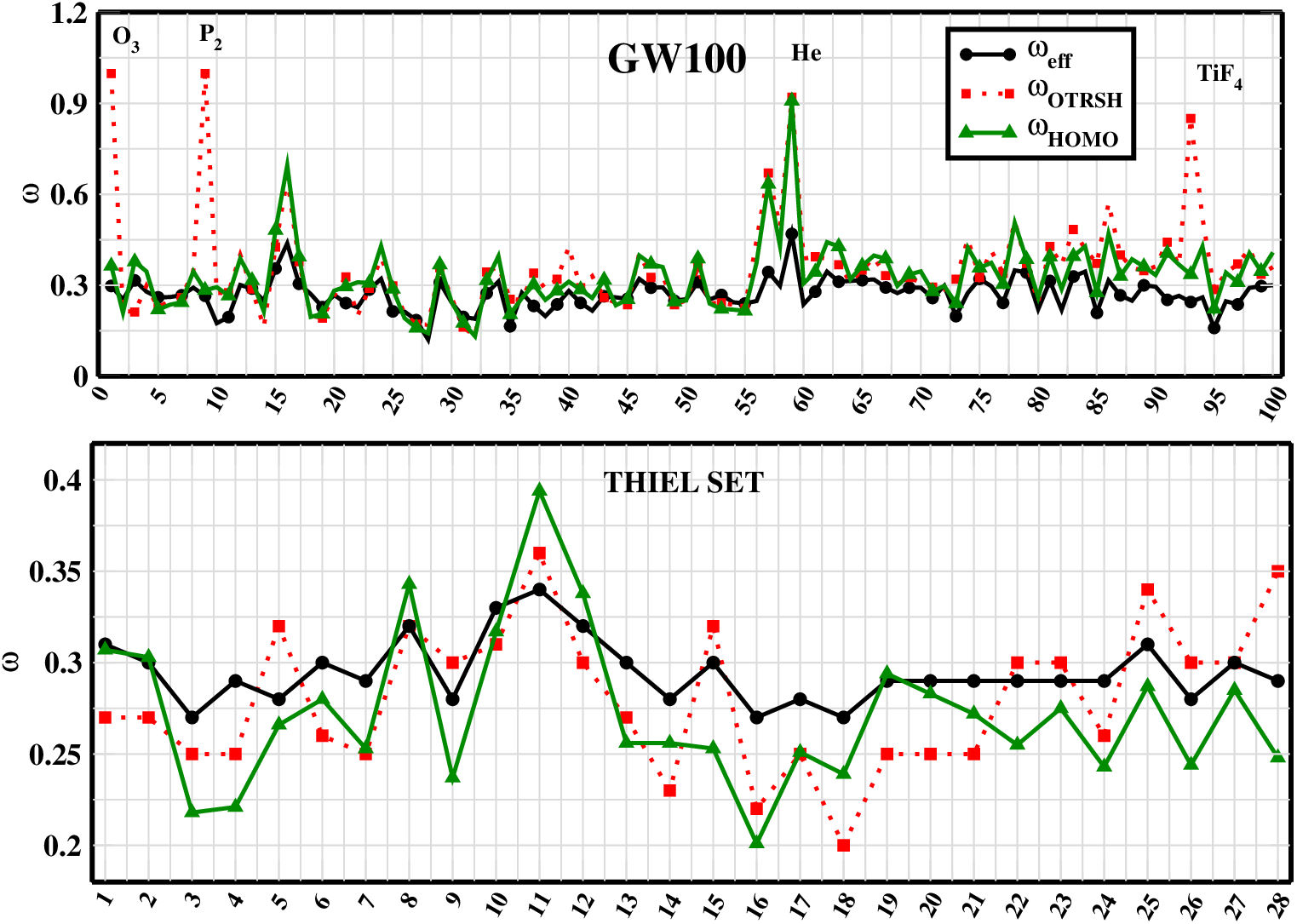}    
    \caption{
    The figure shows the results of the tuning procedure from Eq.~\ref{eq:omega_form}, with all parameters given in bohr$^{-1}$. Results for the GW100 test set are in the top panel, and those for Thiel's set are in the bottom panel. The numbering on the x-axis aligns with the system order listed in Tables S1-S3 of the SI\cite{supplementary}.}
    
 \label{omega_gw100_bse}
\end{figure}

\begin{figure}
    \centering    \includegraphics[width=0.9\columnwidth]{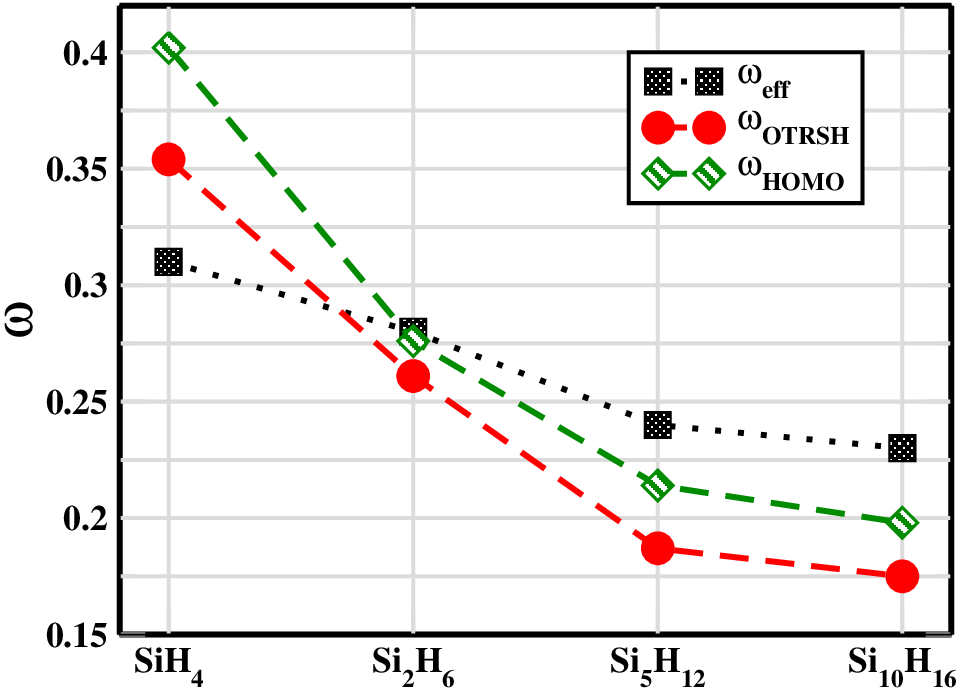}    \caption{
    The range separation parameters $\mathrm{\omega}$ for few hydrogenated silicon quantum dots (Si$_n$H$_m$). The complete data is available in Table S4 of the SI\cite{supplementary}.}
    \label{silicon_plot}
\end{figure}

\begin{figure*}
    \centering    \includegraphics[width=1\textwidth]{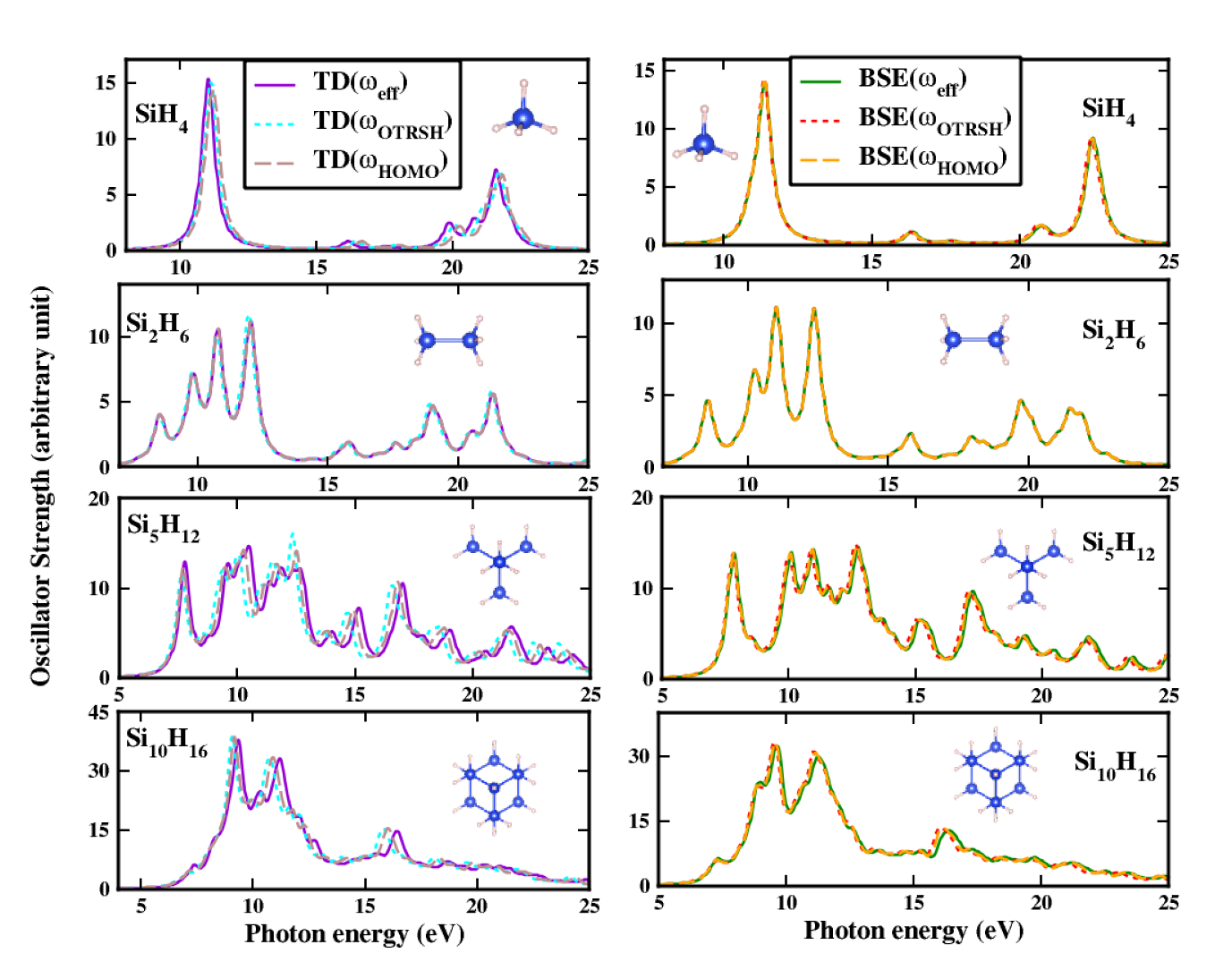}    
   \caption{A comparison of the photoabsorption spectra for silicon clusters calculated with TDDFT (left panel) and the many-body BSE (right panel). The simulations were performed with the LC-$\mathrm{\omega}$PBEh functional, considering all RS tuning parameters. The calculations have been performed with TZVP basis set.}
    \label{photoabsorption}
\end{figure*}

\begin{table}
\centering
\caption{
Mean signed error (MSE), mean absolute error (MAE), and mean absolute relative error (MARE) for IPs of the GW100 set obtained from GKS and G$_0$W$_0$ HOMO energies. GW100 results are referenced to CCSD(T) values. Full numerical results are available in the SI~\cite{supplementary}.}
\label{tab:gw100_errors1}
\renewcommand{\arraystretch}{1.2}
\setlength{\tabcolsep}{9pt}
\resizebox{0.4\paperwidth}{!}{
\begin{tabular}{lccc}
\hline\hline
\textbf{Tuning} & \textbf{MSE (eV)} & \textbf{MAE (eV)} & \textbf{MARE (\%)} \\
\hline
\multicolumn{4}{c}{DFT (GKS)} \\
\hline
$\omega_{\mathrm{eff}}$   &  0.305 & 0.367 & 3.064 \\
$\omega_{\mathrm{OTRSH}}$   & -0.127 & 0.226 & 1.931 \\
$\omega_{\mathrm{HOMO}}$  & -0.051 & 0.124 & 1.156 \\[2pt]
\hline
\multicolumn{4}{c}{MBPT ($\mathrm{G}_0\mathrm{W}_0@\mathrm{GKS}$)} \\
\hline
$\omega_{\mathrm{eff}}$ &  -0.008 & 0.122 & 1.174 \\
$\omega_{\mathrm{OTRSH}}$   & -0.044 & 0.139 & 1.300 \\
$\omega_{\mathrm{HOMO}}$ & -0.037 & 0.137 & 1.298 \\[3pt]
\hline\hline
\end{tabular}}
\end{table}

\begin{table}
\centering
\caption{
Mean signed error (MSE), mean absolute error (MAE), and mean absolute relative error (MARE) for singlet (S), triplet (T) excitation energies of Thiel's test set. The Thiel's excitation energies are compared with the theoretical best estimates~\cite{bte-ref} results. Full numerical results are available in the SI~\cite{supplementary}.}
\label{tab:gw100_errors2}
\renewcommand{\arraystretch}{1.2}
\setlength{\tabcolsep}{9pt}
\resizebox{0.4\paperwidth}{!}{
\begin{tabular}{lccc}
\hline\hline
\textbf{Method} & \textbf{MSE (eV)} & \textbf{MAE (eV)} & \textbf{MARE (\%)} \\
\hline
\multicolumn{4}{c}{\textbf{{TDDFT ($@$GKS)}}} \\
\hline
\multicolumn{4}{c}{{Singlet}} \\
\hline
$\omega_{\mathrm{eff}}$ & -0.27 & 0.34 &  6.30 \\
$\omega_{\mathrm{OTRSH}}$  & -0.28 & 0.39 &  7.33 \\
$\omega_{\mathrm{HOMO}}$  &  -0.22& 0.31 & 5.70  \\
\hline
\multicolumn{4}{c}{{Triplet}} \\
\hline
$\omega_{\mathrm{eff}}$ &  0.43 & 0.49 & 12.14 \\
$\omega_{\mathrm{OTRSH}}$ &  0.55 & 0.64 & 15.47 \\
$\omega_{\mathrm{HOMO}}$  & 0.41  &  0.49& 11.76 \\
\hline
\multicolumn{4}{c}{{Singlet + Triplet}} \\
\hline
$\omega_{\mathrm{eff}}$         &  0.08 & 0.42 &  9.22 \\[2pt]
$\omega_{\mathrm{OTRSH}}$         &  0.14 & 0.52 & 11.40 \\[2pt]
$\omega_{\mathrm{HOMO}}$        &0.10   &  0.40&  8.73\\
\hline
\multicolumn{4}{c}{\textbf{{MBPT} (BSE$@G_0W_0@$GKS)}} \\
\hline
\multicolumn{4}{c}{{Singlet}} \\
\hline
$\omega_{\mathrm{eff}}$ & -0.15 & 0.22 &  4.17 \\
$\omega_{\mathrm{OTRSH}}$  & -0.14 & 0.23 &  4.30 \\
$\omega_{\mathrm{HOMO}}$  &-0.11  &0.22  &  4.09 \\
\hline
\multicolumn{4}{c}{{Triplet}} \\
\hline
$\omega_{\mathrm{eff}}$ &  0.30 & 0.40 &  9.25 \\
$\omega_{\mathrm{OTRSH}}$  &  0.20 & 0.41 & 10.28 \\
$\omega_{\mathrm{HOMO}}$ & 0.21  & 0.40 &  9.65\\
\hline
\multicolumn{4}{c}{{Singlet + Triplet}} \\
\hline
$\omega_{\mathrm{eff}}$         &  0.08 & 0.31 &  6.71 \\[2pt]
$\omega_{\mathrm{OTRSH}}$         &  0.03 & 0.32 &  7.29 \\[2pt]
$\omega_{\mathrm{HOMO}}$         &  0.05 & 0.31 &  6.87 \\
\hline\hline
\end{tabular}}
\end{table}

\section{Results and Discussions}
\label{sec:results}

\subsection{Comparison of $\mathrm{\omega}$}
First, we examine the trends in the tuned range-separation parameters $\omega_{\mathrm{OTRSH}}$, $\omega_{\mathrm{HOMO}}$, and $\omega_{\mathrm{eff}}$ defined by 
Eq.~\ref{eq:omega_form}, for GW100 set and Thiel's test set. A comparison among these values is presented in Fig.~\ref{omega_gw100_bse}. While the three definitions yield broadly consistent results, notable deviations are observed for certain molecules in the GW100 set, particularly between $\omega_{\mathrm{OTRSH}}$ (even $\omega_{\mathrm{HOMO}}$) and $\omega_{\mathrm{eff}}$. Specifically, $\omega_{\mathrm{OTRSH}}$ spans the range $[0.13, 1.00]$ bohr$^{-1}$ and $\omega_{\mathrm{HOMO}}$ lies within $[0.13, 0.91]$ bohr$^{-1}$, while $\omega_{\mathrm{eff}}$ remains more narrowly distributed in the range $[0.12, 0.47]$ bohr$^{-1}$, well within the empirically optimal window of most of the RSH functionals~\cite{RohrMartHerb2009}.
The effective scheme is also not sensitive to the choice of the reference density used to calculate the  $\omega_{\mathrm{eff}}$ value. This is shown in Table S15 of the SI where we report the latter for PBE, B3LYP and CCSD densities.  The results indicate that $\omega_{\mathrm{eff}}$ is nearly independent of the choice of method and therefore only weakly sensitive to it (see the \% offsets calculated with respect to the PBE density used in the main text). For Ne and He, the maximum difference between the computed $\omega_{\mathrm{eff}}$ values is approximately 2\%
(see Table S15 in SI file). These results support the statement in Ref.\cite{singh2025simplifiedphysicallymotivateduniversally} that the proposed scheme acts as an averaging procedure; thus, variations in the input density across different electronic-structure methods do not significantly affect the resulting $\omega_{\mathrm{eff}}$ values.
%

In general, the $\omega_{\mathrm{eff}}$ values follow a trend similar to that observed for $\omega_{\mathrm{OTRSH}}$ and $\omega_{\mathrm{HOMO}}$, except for a few cases. 
 For certain molecules, $\omega_{\mathrm{OTRSH}}$ reaches unusually large values (0.90-0.99 bohr$^{-1}$), as observed for 10028-15-6, 12185-09-0, and 7783-63-3, corresponding to O$_3$, P$_2$, and TiF$_4$, respectively. These cases are characterized by either nearly degenerate frontier orbitals or unbound anionic states, both of which are well known to hinder IP- or HOMO-based tuning schemes~\cite{ot-rsh-gw100}. In contrast, $\omega_{\mathrm{eff}}$ provides physically reasonable values for these cases: 0.298 bohr$^{-1}$ for O$_3$, 0.265 bohr$^{-1}$ for P$_2$, 
 and 0.245 bohr$^{-1}$ for TiF$_4$. 


For systems in Thiel's set, the resulting values of $\omega_{\mathrm{eff}}$ 
gives stable behavior following a trend similar to $\omega_{\mathrm{OTRSH}}$ and $\omega_{\mathrm{HOMO}}$, except for a few cases, where the deviation is larger. Interestingly, for most molecules, the corresponding $\omega_{\mathrm{eff}}$ average values are found to be approximately 0.3~bohr$^{-1}$, which is consistent with the optimal range expected for LC-type functionals. This further supports the robustness and general applicability of the effective approach.

\subsection{Ionization Potential}

First, we analyze the performance of various tuning schemes against the GW100 reference IPs, using both the GKS and G$_0$W$_0$ HOMO orbital energies. Accurate assessment of these IPs is a crucial first step, as the GKS and subsequently G$_0$W$_0$ corrected eigenvalues provide the essential input for the following BSE calculations.

To this end, in Table~\ref{tab:gw100_errors1}, we compare
the impact of tuning scheme on GKS 
and the corresponding G$_0$W$_0$ results. Table~\ref{tab:gw100_errors1} summarizes the mean signed errors (MSEs), mean absolute errors (MAEs), and mean absolute relative errors (MAREs) of these predictions computed with respect high-level CCSD(T) reference results. 
Full results are provided in the SI file~\cite{supplementary}, including the errors computed relative to experimental values. The corresponding data for the def2-QZVP basis set~\cite{def2-family} are also reported in the SI~\cite{supplementary} Tables S9 and S10; the overall trend remains essentially unchanged upon moving to the larger basis set. 
Additionally, Tables S7 and S8 present results obtained with various functionals as starting points for G$_0$W$_0$ calculations on the GW100 test set, further confirming that RSH constitutes the optimal choice for such an analysis.

One can note that choice of tuning parameter strongly affects the KS-DFT (GKS) results, whereas its impact becomes much weaker after applying  G$_0$W$_0$ corrections.
At the KS-DFT (GKS) level, \(\omega_{\mathrm{eff}}\) produces the largest and clearly systematic overestimation of the IPs (MSE \(=0.305\) eV), together with comparatively large absolute errors (MAE \(=0.367\) eV; MARE \(=3.064\%\)). 
This is not surprising, since the \(\omega_{\mathrm{eff}}\) tuning scheme does not aim to reproduce any particular orbital energy.
In contrast, \(\omega_{\mathrm{OTRSH}}\) substantially reduces the bias and improves the overall accuracy (MSE \(=-0.127\) eV; MAE \(=0.226\) eV; MARE \(=1.931\%\)). The best KS-DFT performance is achieved with $\omega_{\mathrm{HOMO}}$, as expected given its more sophisticated tuning procedure, yielding a near-zero mean signed error (MSE $=-0.051$ eV) and the smallest overall errors (MAE $=0.124$ eV; MARE $=1.156\%$).\\
Upon moving to MBPT (G$_0$W$0$@GKS), the $\omega{_{\mathrm{eff}}}$ starting point yields the best overall performance, with an MSE \(=-0.008\) eV, an MAE \(=0.122\) eV, and an MARE \(=1.174\%\). By contrast, the other two tuning schemes $\omega{_{\mathrm{OTRSH}}}$ and $\omega{_{\mathrm{HOMO}}}$  perform nearly identically, with MSEs (\(-0.044\) and  \(-0.037\) eV), MAEs ($0.139$ and $0.137$ eV), and MAREs ($1.300$\% and $1.298$\%), respectively.
Notably, G$_0$W$_0$ dramatically improves the \(\omega_{\mathrm{eff}}\) starting point (MAE reduced from \(0.367\) to \(0.122\) eV), indicating that the GW correction largely compensates for the KS-level overestimation. Meanwhile, because $\omega_{\mathrm{HOMO}}$ already provides excellent accuracy at the KS-DFT level, the subsequent G$_0$W$_0$ correction results in only minor changes to the overall statistical performance. In principle, the $\omega_{\mathrm{HOMO}}$ scheme determines the range-separation parameter by enforcing equality between the GKS and G$_0$W$_0$ HOMO energies. In practice, however, the optimal $\omega$ is determined numerically by spline interpolation of calculations performed on a finite grid of sampled $\omega$ values. Consequently, this condition is satisfied only approximately, leading to the small residual differences observed in the reported MSE, MAE, and MARE values.
However, we recall that to obtain the $\omega_{\mathrm{HOMO}}$ parameter values one requires to perform a complex, multi-step optimization process designed to align the HOMO energies from G$_0$W$_0$@GKS($\omega$) and GKS$(\omega)$ calculations. For this reason the $\omega_{\mathrm{HOMO}}$ based IP G$_0$W$_0$ results are far more expensive to obtain in comparison to effective tuning counterpart. 
Thus, the close agreement of all G$_0$W$_0$ results underscores the effectiveness of the $\omega_{\mathrm{eff}}$ scheme as a transferable black-box alternative to the more expensive optimally tuned RSHs.\\

\subsection{Excitation Energies}

As a next step, we analyze the excitation energies within the TDDFT and BSE frameworks. Table~\ref{tab:gw100_errors2} summarizes the error statistics for Thiel's test set~\cite{bte-ref}, including both singlet and triplet excitations, where the reference values correspond to the theoretical best estimates (TBE) from Ref.~\cite{ot-rsh-gw100}. Within the MBPT framework, the BSE@G$_0$W$_0$@GKS calculations starting from $\omega_{\mathrm{eff}}$ based reference perform comparably to those based on the more elaborate $\omega_{\mathrm{OTRSH}}$ and $\omega_{\mathrm{HOMO}}$ counterparts. For singlet excitations, the errors obtained for BSE with $\omega_{\mathrm{eff}}$ (MAE = 0.22 eV, MARE = 4.17\%) are nearly identical to those with $\omega_{\mathrm{OTRSH}}$ (MAE = 0.23 eV, MARE = 4.30\%), and $\omega_{\mathrm{HOMO}}$ (MAE = 0.22 eV, MARE = 4.09\%), (see Table S11-S12 in SI).
Triplet excitation are much sensitive to different tuning protocols, overall the trend remains similar. Here the $\omega_{\mathrm{eff}}$ results yeild MAE = 0.40 eV and MARE = 9.25\%) which remain very close to those for $\omega_{\mathrm{OTRSH}}$ (MAE = 0.41 eV, MARE = 10.28\%) and $\omega_{\mathrm{HOMO}}$ (MAE = 0.40 eV, MARE = 9.65\%) counterparts(see Table S13-S14 in SI). 


It is noteworthy that $\omega_{\mathrm{OTRSH}}$ has previously been identified as one of the most accurate starting points for BSE@G$_0$W$_0$@GKS calculations among a broad range of DFAs~\cite{ot-rsh-gw100}. Other commonly used functionals, such as PBE, PBE0, B3LYP, CAM-B3LYP, and BHLYP, generally yield significantly larger MAEs when employed as starting points for BSE@G$_0$W$_0$@GKS~\cite{ot-rsh-gw100}. We refer to Table S16 of the SI for the error statistics for other functionals. Therefore, the comparable or even slightly improved performance of the computationally efficient $\omega_{\mathrm{eff}}$ method highlights its potential as a practical and reliable alternative for generating accurate excitation spectra.

Secondly, in Table~\ref{tab:gw100_errors2} we present a comparison of TDDFT excitation energies calculated using different variants of the tuning parameter. 
Our previous work~\cite{singh2025simplifiedphysicallymotivateduniversally} showed that TDDFT($\omega_{\mathrm{eff}}$) provides enhanced performance for CT excitations and optical gap predictions in organic photovoltaic (OPV) systems. This advantage persists in the present analysis of Thiel's test set, where both singlet and triplet excitation energies predicted by TDDFT($\omega_{\mathrm{eff}}$) exhibit similar performance to TDDFT($\omega_{\mathrm{HOMO}}$), measured with respect to TBEs outperforming the TDDFT($\omega_{\mathrm{OTRSH}}$) results. 

Moreover, the comparison with traditional DFAs commonly employed in TDDFT calculations~\cite{ot-rsh-gw100} generally produces substantially larger MAEs than either of the above tuning variants, reinforcing the importance of incorporating range separation and long-range exchange effects.

Nonetheless, one can see that the adoption of a physically motivated and analytically simple effective range-separation parameter, $\omega_{\mathrm{eff}}$, within the RSH framework not only maintains high predictive accuracy but also significantly reduces the computational overhead associated with optimal-tuning procedures. Consequently, the $\omega_{\mathrm{eff}}$ based RSH approach provides a robust and efficient foundation for both TDDFT and BSE@G$_0$W$_0$ calculations of excitation energies across diverse molecular systems.

\begin{table*}
\centering
\caption{HOMO, HOMO-LUMO gaps, optical gaps, and exciton binding energies using different tuning protocols for TDDFT and MBPT variants are presented. All energies are in eV. Diffusion quantum Monte Carlo (DMC) results from ref.~\cite{Benedict2003Calculation, DMC-PRL}}
\label{tab:si_cluster}
\begin{tabular}{llccccccc}
\hline\hline
\multirow{2}{*}{\textbf{Cluster}} & \multirow{2}{*}{\textbf{Property}} & \multicolumn{3}{c}{\textbf{{DFT/TDDFT}}} & \multicolumn{3}{c}{\textbf{G$_0$W$_0$/BSE}} & \multirow{2}{*}{\textbf{DMC}} \\
\cline{3-8}
 & & $\omega_{\mathrm{eff}}$ & $\omega_{\mathrm{OTRSH}}$ & $\omega_{\mathrm{HOMO}}$ & $\omega_{\mathrm{eff}}$ & $\omega_{\mathrm{OTRSH}}$ & $\omega_{\mathrm{HOMO}}$ & \\
\hline
\multirow{5}{*}{SiH$_4$} & HOMO & -12.50 & -13.11 & -12.97 & -12.71 & -12.77 & -12.75 & -- \\
 & E$_{\text{HOMO-LUMO}}$ & 15.72 & 16.56 & 16.37 & 16.45 & 16.53 & 16.51 & -- \\
 & E$_{\text{ex}}$ & 6.11 & 6.64 & 6.53 & 6.56 & 6.54 & 6.54 & -- \\
 & E$_{\text{opt}}$ & 9.61 & 9.92 & 9.84 & 9.89 & 9.99 & 9.97 & 9.2 \\
\hline
\multirow{5}{*}{Si$_2$H$_6$} & HOMO & -10.66 & -10.98 & -10.66 & -10.39 & -10.43 & -10.39 & -- \\
 & E$_{\text{HOMO-LUMO}}$ & 13.04 & 13.54 & 13.04 & 13.30 & 13.37 & 13.30 & -- \\
 & E$_{\text{ex}}$ & 5.29 & 5.67 & 5.29 & 5.51 & 5.53 & 5.51 & -- \\
 & E$_{\text{opt}}$ & 7.75 & 7.87 & 7.75 & 7.79 & 7.84 & 7.79 & 7.5 \\
\hline
\multirow{5}{*}{Si$_5$H$_{12}$} & HOMO & -10.07 & -10.15 & -9.90 & -9.67 & -9.68 & -9.64 & -- \\
 & E$_{\text{HOMO-LUMO}}$ & 11.61 & 11.75 & 11.34 & 11.67 & 11.69 & 11.63 & -- \\
 & E$_{\text{ex}}$ & 4.73 & 4.84 & 4.52 & 4.70 & 4.71 & 4.69 & -- \\
 & E$_{\text{opt}}$ & 6.88 & 6.91 & 6.82 & 6.97 & 6.98 & 6.94 & 6.8 \\
\hline
\multirow{5}{*}{Si$_{10}$H$_{16}$} & HOMO & -9.80 & -9.75 & -9.53 & -9.15 & -9.15 & -9.11 & -- \\
 & E$_{\text{HOMO-LUMO}}$ & 10.31 & 10.24 & 9.89 & 9.97 & 9.96 & 9.89 & -- \\
 & E$_{\text{ex}}$ & 4.13 & 4.08 & 3.84 & 3.86 & 3.86 & 3.84 & -- \\
 & E$_{\text{opt}}$ & 6.18 & 6.16 & 6.05 & 6.11 & 6.10 & 6.05 & 6.0 \\
\hline\hline
\end{tabular}
\end{table*}

\begin{table}[h!]
\centering
\scriptsize
\setlength{\tabcolsep}{3pt} 
\caption{Optical absorption energies (eV) for SiH$_4$ and Si$_2$H$_6$. Experimental (Expt) values from Ref.~\cite{jcp-expt}.}
\label{tab:sih}
\begin{tabular}{lccccccc}
\hline\hline
 & & \multicolumn{3}{c}{TDDFT(@GKS)} & \multicolumn{3}{c}{BSE@G$_0$W$_0$@GKS} \\
\cline{3-8}
Cluster & Expt & $\omega_{\mathrm{eff}}$ & $\omega_{\mathrm{OTRSH}}$ & $\omega_{\mathrm{HOMO}}$ 
        & $\omega_{\mathrm{eff}}$ & $\omega_{\mathrm{OTRSH}}$ & $\omega_{\mathrm{HOMO}}$ \\
\hline
\multirow{3}{*}{SiH$_4$} 
 & 8.80 & 9.61 & 9.92 & 9.84 & 9.89 & 9.99 & 9.97 \\
 & 9.70 & 9.85 & 10.17 & 10.09 & 10.16 & 10.26 & 10.23 \\
 & 10.70 & 10.70 & 10.82 & 11.02 & 10.80 & 10.88 & 10.86 \\
\hline
\multirow{4}{*}{Si$_2$H$_6$} 
 & 7.60 & 7.75 & 7.87 & 7.75 & 7.79 & 7.85 & 7.60 \\
 & 8.40 & 8.55 & 8.72 & 8.55 & 8.58 & 8.63 & 8.40 \\
 & 9.50 & 9.29 & 9.51 & 9.29 & 9.64 & 9.71 & 9.50 \\
 & 9.90 & 9.67 & 10.01 & 9.67 & 9.93 & 10.00 & 9.90 \\
\hline\hline
\end{tabular}
\end{table}
\subsection{Hydrogenated silicon quantum dots}

Going beyond molecular systems, the effective range-separation parameter $\omega_{\mathrm{eff}}$ is also highly relevant for size-dependent nanocrystalline materials. A prototypical example is the family of hydrogenated silicon quantum dots (Si$_n$H$_m$), where quantum confinement plays a crucial role in determining their electronic and optical properties. These systems have been extensively studied using a range of theoretical approaches, including DFT, TDDFT, the GW approximation, and BSE~\cite{Ogut1999Optical,Benedict2003Calculation,Ogut1997Quantum,expt-prl,PRL2000Si,Vasiliev2001abinitio,Puzder2002Surface,DMC-PRL}. All these studies report strong confinement-induced blue-shifts in the optical gaps and sizable exciton binding energies~\cite{Ogut1999Optical,Ogut1997Quantum}. Due to the lack of experimental benchmarks for these quantum dots, theoretical predictions from high-level MBPT methods such as GW and BSE~\cite{Ogut1999Optical,Benedict2003Calculation,expt-prl}, as well as from diffusion Monte Carlo (DMC)~\cite{Benedict2003Calculation}, serve as essential references.

In Fig.~\ref{silicon_plot}, we present the size-dependent variation of $\omega_{\mathrm{eff}}$ for a series of Si$_n$H$_m$ clusters. Interestingly, $\omega_{\mathrm{eff}}$ shows a smooth and physically significant trend with respect to cluster size, consistent with previously observed trends~\cite{singh2025simplifiedphysicallymotivateduniversally} and closely tracks the behavior of the IP-based tuning parameter $\omega_{\mathrm{OTRSH}}$~\cite{SteiEiseHele2010} and $\omega_{\mathrm{HOMO}}$. The fact that $\omega_{\mathrm{eff}}$ nearly approximates $\omega_{\mathrm{HOMO}}$ (and $\omega_{\mathrm{OTRSH}}$), even for relatively large systems, underscores its practical utility. The direct evaluation of $\omega_{\mathrm{HOMO}}$ (and $\omega_{\mathrm{OTRSH}}$) becomes computationally prohibitive for large clusters, while $\omega_{\mathrm{eff}}$ offers a viable and less demanding alternative.

Table~\ref{tab:si_cluster} presents the calculated HOMO,
HOMO-LUMO gaps, optical gaps, and exciton binding energies obtained from different tuning protocols. The results from both TDDFT and MBPT are internally consistent. With increasing cluster size, the HOMO becomes less negative and both the HOMO-LUMO gap and the optical gap E$_{\mathrm{opt}}$ decrease. In parallel, the exciton binding energy E$_{\mathrm{ex}}$ decreases systematically (from $\sim 6$~eV to $\sim 4$~eV), consistent with weaker electron-hole binding and enhanced screening in larger clusters. At the TDDFT level, the tuning protocol introduces a noticeable spread in gaps and excitation energies. Overall, $\omega_{\mathrm{OTRSH}}$ tends to yield slightly larger HOMO-LUMO gaps and E$_{\mathrm{opt}}$ than $\omega_{\mathrm{HOMO}}$, while $\omega_{\mathrm{eff}}$ typically provides intermediate values. This behavior suggests that $\omega_{\mathrm{eff}}$ balances the effective range of exchange in a way that moderates both quasiparticle-like gaps and excitonic effects within TDDFT. In contrast, MBPT results are only weakly dependent on the underlying tuning choice, including $\omega_{\mathrm{eff}}$, with variations typically limited to a few hundredths to at most $\sim 0.1$~eV. This indicates that quasiparticle corrections (G$_0$W$_0$) and the explicit electron-hole interaction (BSE) largely mitigate starting-point differences introduced by the tuned functional. Comparison with DMC optical gaps shows that TDDFT and BSE@G$_0$W$_0$ generally overestimate E$_{\mathrm{opt}}$ for the smallest cluster, but the discrepancy decreases with size; for Si$_{10}$H$_{16}$ the calculated optical gaps approach the DMC reference. Overall, Table~\ref{tab:si_cluster} suggests that tuning primarily affects TDDFT predictions, whereas BSE@G$_0$W$_0$ provides more robust, less protocol-dependent optical gaps, and that $\omega_{\mathrm{eff}}$ behaves as a stable tuning strategy.

Expanding the analysis, Table~\ref{tab:sih} reports the optical absorption energies of SiH$_4$ and Si$_2$H$_6$, two systems for which reliable experimental reference data are available. For both clusters, the TDDFT results obtained with the $\omega_{\mathrm{eff}}$ parameter are consistently more accurate, showing closer agreement with experiment than those computed using $\omega_{\mathrm{OTRSH}}$ or $\omega_{\mathrm{HOMO}}$. The same trend holds for the BSE calculations: although BSE generally overestimates excitation energies, the overestimation is substantially reduced when $\omega_{\mathrm{eff}}$ is used in place of $\omega_{\mathrm{OTRSH}}$ or $\omega_{\mathrm{HOMO}}$.

Figure~\ref{photoabsorption} compares the corresponding optical spectra from TDDFT (left) and BSE (right) using the three tuning protocols. In TDDFT (left panel), the spectra differ only modestly, with noticeable changes limited to a subset of features. By contrast, the BSE spectra (right panel) display a clear, peak-by-peak correspondence across the three tuning choices.

Overall, these results show that $\omega_{\mathrm{eff}}$ not only captures the correct size dependence of key observables, but also provides a computationally efficient and physically well-motivated choice for describing electronic excitations in nanostructured materials.

\section{Conclusion}
\label{sec:conclusion}
In this work, we have systematically assessed the performance of a recently proposed, physically motivated, and computationally efficient range-separation tuning strategy for RSH functionals in the context of MBPT methods. Specifically, we demonstrated that the effective screening parameter $\omega_{\mathrm{eff}}$, derived from the average electron density, serves as a robust, transferable and much computationally accessible alternative to conventional tuning schemes based on the gap tuning or HOMO alignment.

By employing $\omega_{\mathrm{eff}}$ within an RSH framework as the starting point for single-shot G$_0$W$_0$ and BSE calculations, we achieved ionization potentials and excitation energies in excellent agreement with high-level reference data for the GW100 set and Thiel's benchmark test set. Notably, the accuracy of G$_0$W$_0$@GKS($\omega_{\mathrm{eff}}$) matches that of optimally tuned functionals, yet it completely avoids their demanding, multi step tuning procedure. Although the $\omega_{\mathrm{HOMO}}$ approach may appear slightly more accurate in some cases, this comes again with a substantial computational burden.
Furthermore, for a series of hydrogenated silicon nanoclusters (Si$_n$H$_m$), we showed that $\omega_{\mathrm{eff}}$ captures the expected size-dependent trends and yields optical gaps and exciton binding energies consistent with those from diffusion Monte Carlo and BSE calculations. This highlights the applicability of $\omega_{\mathrm{eff}}$ beyond molecular systems, making it a practical tool for exploring excited-state properties in nanoscale and extended materials.

Overall, our results indicate that GKS($\omega_{\mathrm{eff}}$) serves as a physically justified, black-box replacement for standard tuning methods, offering a dependable initial setup for G$_0$W$_0$ and BSE computations while considerably lowering the computational cost. This method is especially advantageous for large-scale and high-throughput investigations and simulations with molecules within different solvent systems, in which conventional tuning approaches become impractical. 

Notably, the effective tuning strategy has recently been adopted to predict CT processes more accurately than wavefunction theory, supporting the design of efficient dye-sensitized solar cells based on organic dyes.\cite{doping}. Similar applications of the effective-tuning protocol are being actively pursued in our group, and the corresponding results will be reported in due course.

\section*{Acknowledgements}
S.\'S. acknowledges the financial support from the National Science Centre, Poland (grant no. 2021/42/E/ST4/00096). 

\section*{Supporting Information}
The Supporting Information is available free of charge at \cite{supplementary}

\begin{itemize}
  \item Values of the range-separation parameters $\omega_{\mathrm{eff}}$, $\omega_{\mathrm{OTRSH}}$, and $\omega_{\mathrm{HOMO}}$ for GW100, Thiel set and silicon cluster and all raw data analyzed in the work.
  
  \item The optimized molecular geometries of all Silicon clusters are available via the following public GitHub repository:
\small{\texttt{https://github.com/aditisingh4812/Silicon-cluster}}
\end{itemize}



\twocolumngrid

\bibliography{reference}

\end{document}